\documentclass[conference]{IEEEtran}
\IEEEoverridecommandlockouts
\usepackage{cite}
\usepackage{amsmath,amssymb,amsfonts}
\usepackage{algorithmic}
\usepackage{graphicx}
\usepackage{textcomp}
\usepackage{xcolor}
\usepackage{multirow}
\usepackage{soul}
\usepackage{url}
\usepackage{threeparttable}
\usepackage{comment}
\usepackage{hyperref}
\hypersetup{hidelinks,
	colorlinks=true,
	allcolors=black,
	pdfstartview=Fit,
	breaklinks=true}

\def\BibTeX{{\rm B\kern-.05em{\sc i\kern-.025em b}\kern-.08em
    T\kern-.1667em\lower.7ex\hbox{E}\kern-.125emX}}
\begin{document}

\setlength{\textfloatsep}{4pt plus 0.3pt minus 0.3pt}
\setlength{\intextsep}{4pt plus 0.3pt minus 0.3pt}

\title{SRAM-PG: Power Delivery Network Benchmarks from SRAM Circuits \\
\thanks{This work is partially supported by the National
Natural Science Foundation of China (No. 62204141), and Beijing Natural Science Foundation (No. Z230002). S. Shen and Z. Liu contributed equally to this work. W. Yu is the corresponding author.}
}

\author{\IEEEauthorblockN{Shan Shen, Zhiqiang Liu, Wenjian Yu}
\IEEEauthorblockA{\textit{Dept. Computer Science \& Tech., BNRist, Tsinghua Univ., Beijing, China} \\
Email: shanshen@tsinghua.edu.cn, liu-zq20@mails.tsinghua.edu.cn, yu-wj@tsinghua.edu.cn
}}

\maketitle

\begin{abstract}
Designing the power delivery network (PDN) in very large-scale integrated (VLSI) circuits is increasingly important, especially for nowadays low-power integrated circuit (IC) design. In order to ensure that the designed PDN enables a low level of voltage drop and noise which is required for the success of IC design, accurate analysis of PDN is largely demanded and brings a challenge of computation during the whole process of IC design. This promotes the research of efficient and scalable simulation methods for PDN.
However, the lack of sufficient public PDN benchmarks hinders the relevant research. 
To this end, we construct and release a set of PDN benchmarks (named \emph{SRAM-PG}) from SRAM circuit design in this work. The benchmarks are obtained from realistic and state-of-the-art SRAM designs, following a workflow for generating the post-layout PDN netlists with full RC parasitics. With careful modeling of load currents, the benchmarks reflect the dynamic work mode of the IC and can be used for both transient and DC analysis. The benchmarks are derived from the designs for diverse applications. And, sharing them in the public domain with detailed descriptions would largely benefit the relevant research. The whole set of benchmarks is available at \href{github}{https://github.com/ShenShan123/SRAM-PG}.
\end{abstract}


\section{Introduction}
With the advancement of process technology, the design and analysis of power delivery networks (PDNs) in integrated circuits (ICs) becomes more and more important. 
A well-designed PDN minimizes voltage fluctuations, reduces power losses, and ensures that each component within the IC receives adequate power. Inefficient power delivery can lead to performance degradation, increased power consumption, and even failure of the IC. Therefore, accurate analysis of PDN is crucial in the design process because it helps to identify the potential power integrity problems, including IR drop (voltage drop), electromigration (EM), the derived thermal issue, etc. For example, the IR drop can cause the circuit to malfunction or fail if the voltage level drops below the minimum required voltage for a component.

A lot of research efforts have been devoted to IR drop analysis or PDN simulation in the past two decades. They include the methods based on direct equation solver \cite{cholmod2008,davis2017suitesparse,powerrush2012,pgtsolver,parsub}, the iterative equation solver\cite{feng2019grass,fegrass,yang2014powerrush,drw,liu2022pursuing,epsim,li2023unleashing,li2023parallel,mst-guided17} and the specific approaches \cite{sun2007parallel,silva2010efficient,macromodel}. 
Among them, the iterative solver-based approaches have gained a lot of attention, due to their scalability to large-scale cases.
Recently, efficient parallel algorithms based on iterative solver were developed which can accomplish DC analysis of PDNs with up to 0.36 billion nodes within twenty minutes on a normal multi-core computer \cite{pgrassiccad,liu2023pgrass}.
However, most of these algorithms were only validated with two old-dated public PDN benchmarks. More up-to-date and realistic benchmarks are required to promote the research of iterative solvers for PDN simulation.


\begin{table*}[tb]
\setlength{\abovecaptionskip}{0 cm}
\begin{center}
\caption{Comparison of Different Open-sourced PDN Benchmarks}\label{tab_comp}
\begin{threeparttable}
\begin{tabular}{l|llllll}
\hline
 & Technology & Scale (Node \#) & Parasitic Type & Analysis Mode & Current Map & Release Date \\ \hline
IBMPG\cite{ibmpg} & N/A & 30.6K$\sim$1.6M & R+C\textsubscript{G}\tnote{2} & .DC + .Tran & Synthesized & 2008 \\
THUPG\cite{thupg} & TSMC65nm & 4.9M$\sim$60.3M & R & .DC & Synthesized & 2012 \\
BeGAN\cite{began} & \begin{tabular}[c]{@{}l@{}}Nangate 45nm\cite{openroad},\\ SkyWater130nm\cite{skywater}, \\ ASAP 7nm\cite{asap7}\end{tabular} & 18.4K$\sim$455.1K\tnote{1} & R & .DC & Synthesized+Real & 2021 \\
SRAM-PG (this work) & TSMC28nm & 76K$\sim$5.9M & C\textsubscript{G}+C\textsubscript{C}+R & .DC + .Tran & Real 
 & 2024 \\ \hline
\end{tabular}
\begin{tablenotes}
        \footnotesize
        \item[1] This data is collected from the git repository of BeGAN at https://github.com/UMN-EDA/BeGAN-benchmarks/tree/master.
        \item[2] R, C\textsubscript{G}, and C\textsubscript{C} represent the wire resistor, the ground capacitor, and the coupling capacitor, respectively.
\end{tablenotes}
\end{threeparttable}
\end{center}
\vspace{-2em}
\end{table*}

Two PDN benchmarks, IBMPG \cite{ibmpg} and THUPG \cite{thupg},  have been widely used as standard test cases. However, they are dated (released over 10 years ago), and unsuitable for evaluating state-of-the-art algorithms for several reasons. 
Firstly, these benchmarks are based on old process technologies. \cite{ibmpg} is based on Al interconnects, assumes 1.8V supply voltages and most via resistances to be zero. \cite{thupg} is a collection of synthetically generated larger cases based on a test chip design following TSMC 65nm technology. Secondly, they hold some unrealistic assumptions to protect IP. For example, the region-wise uniform currents are assumed in \cite{ibmpg}. Thirdly, the parasitic capacitance on interconnect wires is missing, which does not reflect the effect of advanced process technologies. Recently, a set of synthetic PDN benchmarks based on generative adversarial networks (GAN) and transfer learning techniques was proposed \cite{began}, where the load currents are obtained using transfer learning from satellite images of urban areas. However, these benchmarks are synthesized, small-sized (with less than half a million nodes), and only support DC analysis. Notice that these PDN benchmarks are of little diversity, and all are derived from small- or medium-sized digital designs. More diverse PDN benchmarks are highly demanded for the research of PDN analysis and simulation algorithms.

In the current benchmarks, the cases are divided for DC analysis or transient analysis. In practice, transient analysis of PDN is more demanded \cite{liu2023accuracy}. In the time domain, the voltage at all points of the power network fluctuates. This is because different components within an integrated circuit have different functionality, and may be activated or deactivated at the same time. This time-domain behavior can be quite complicated, especially with the introduction of various power reduction techniques such as multi-voltage domains, clock gating, power gating, dynamic voltage, and frequency scaling (DVFS) \cite{lpmanual}. Furthermore, for a complex system that includes software programmable parts, there will be a large dependence of instantaneous power on the actual software program and data being processed \cite{ibmpg}. 

Upon the urgent need for new and more comprehensive benchmarks for PDN analysis research, in this work, we put forward a set of PDN benchmarks (named \textbf{SRAM-PG}) based on 4 state-of-the-art SRAM designs under TSMC 28nm technology. The key features of the proposed benchmarks include:
\begin{itemize}
\item A full RC network is integrated into the benchmark containing wire resistance, ground capacitance, and coupling capacitance. The PDN benchmarks are generated by academic IC design experts, without any IP issues.
\item The load current is accurately modeled by setting the current sources to the measured values collected from post-layout simulations. The current sources are constructed to mimic the circuit component with different on/off states corresponding to different circuit work modes.
\item The benchmarks are derived from the designs for diverse applications, such as a low-power design, in-memory-computing circuits, and a standard memory module generated by a memory compiler.
\item The benchmarks can be used for both DC and transient analyses, and will be shared in the public domain. 
\end{itemize}

\section{Background}
PDN analysis aims to analyze the supply noise and ground bounce of voltage in integrated circuits. 
For DC analysis, the PDN is modeled as a resistive network. It can be formulated as the following system of linear equations for solving $x$:
\begin{equation}
    \label{equ:dcproblem}
	Gx=b ~,
	\end{equation}
where $G$ is the conductance matrix, $x$ and $b$ denote the unknown vector of node voltages and the vector of current sources respectively. 
For transient analysis, the PDN is modeled as an RC network (probably with L elements as well). 
With modified nodal analysis,  the following differential algebra equations (DAEs) are formulated.
    \begin{equation}
    Gx+C\frac{dx}{dt}=b,
    \end{equation}
where $G$ and $C$ are the conductance matrix and the capacitance matrix respectively. $x$ and $b$ denote the vector of node voltages and current sources respectively. The initial node voltages can be obtained by performing a DC analysis. Then with time integration schemes like the backward Euler scheme, the DAEs are converted to a sequence of linear equation systems for the solution at consecutive time points:
    \begin{equation}
    \label{equ:tranproblem}
    (G+\frac{C}{h})x(t+h)=\frac{C}{h}x(t)+b(t+h) ~.
    \end{equation} 
Here, $x(t\!+\!h)$ is to be solved, $h$ is time step. The transient analysis leads to large computational costs, especially for mixed-signal IC design which consists of a large number of circuit nodes. Therefore, an efficient and scalable PDN simulation algorithm is desired.

The challenges of efficient PDN simulation necessitate the importance of public PDN benchmarks. However, the existing benchmarks are unable to meet the demand.
In IBMPG \cite{ibmpg}, the power grid is only composed of an orthogonal mesh of M1 and M2 layers, which is a somewhat idealized topology. 
Whereas the mesh is not complete (i.e. some wires may be missing or truncated) in a realistic design due to the area constraints. Also, the periodicity and density of the wires may vary because of the different areas of the chip requiring different power consumption. 
It assumes that the connection between the circuit device and the power node occurs at the lowest metal level and only resides at the intersection position. However, at some points, the circuit device may directly connect to a higher metal layer through multiple stacked vias to avoid routing congestion. Moreover, IBMPG ignores the via resistance to reduce the net size. 
In IBMPG, the current maps (CMs) are generated by first assigning a total power for the design, $P_{tot}$, and then randomly distributing it to $N_x\times N_y$ regions of the whole design. In order to realize this, the random weights $\{w_{ij}\}$ are generated to satisfy $\sum^{N_x}_{i=1}\sum^{N_y}_{j=1}w_{ij}=1$. Then, the total current in each region equals $I_{ij}=w_{ij}P_{tot}/V_{DD}$, and a current source with value $I_{ij}/n_{ij}$ is applied to each power node-circuit connection, which means the power is uniformly distributed across the $n_{ij}$ nodes in the region ($i,j$).

THUPG \cite{thupg} is another collection of benchmarks with standard SPICE format compatible with IBMPG. They are extended from a smaller PDN case which is drawn from a test chip design using TSMC 65nm technology. The sizes of these benchmarks are larger than IBMPG benchmarks. However, the voltage drops at some grid nodes are not in a reasonable range because the PDNs are synthesized based on a small design. Another drawback of THUPG is that it is only for the DC analysis. Although the authors in \cite{liu2022pursuing,epsim} modified THUPG benchmarks by adding capacitors and periodical current sources for transient simulation, the resulting benchmarks are still far from practical cases.

The authors of \cite{3dic} proposed a set of PDN benchmarks that reflect the design characteristics of real-world 3D ICs, including variations in die stacking, interconnect length, and the number of power domains. The design cases used by the benchmarks span various sizes and configurations, providing a comprehensive evaluation of the PDNs' performance in 3D ICs. Unfortunately, the benchmarks are not publicly available anymore.

In recent work of BeGAN \cite{began}, the authors leverage generative adversarial networks (GAN) and transfer learning techniques to create PDN benchmarks from a small set of available real circuit data. The BeGAN framework is comprised of two stages: the GAN-based current map generation and the power grid synthesis and power bump assignment using OpeNPDN \cite{openpdn}. In stage 1, the authors first pre-train a GAN using a large set of satellite images of urban regions in a source dataset that has similar characteristics as on-chip CMs. Next, they tune the GAN model using transfer learning (TL) from a source dataset to a target dataset of CMs from a small set of real circuit designs generated by the OpenROAD flow \cite{openroad}. Although BeGAN has generated thousands of benchmarks, they are all for DC analysis and the size of each benchmark is very small (with the number of nodes less than $5\times 10^5$). And, the similarity between the satellite images and the CMs from real design remains questionable. Therefore, BeGAN may be useful for testing some machine-learning-based approaches for early-stage PDN analysis, instead of the accurate PDN simulation algorithms.
Table \ref{tab_comp} lists the open-sourced benchmarks and makes a brief comparison to ours.

\section{Modeling of Power Delivery Network}
In this section, we introduce the overall workflow of generating the benchmarks and show how to build the power delivery network based on the post-layout netlist.

\subsection{Workflow}
\begin{figure}[tb]
  \setlength{\abovecaptionskip}{0 cm}
\centerline{\includegraphics[width=0.9\linewidth]{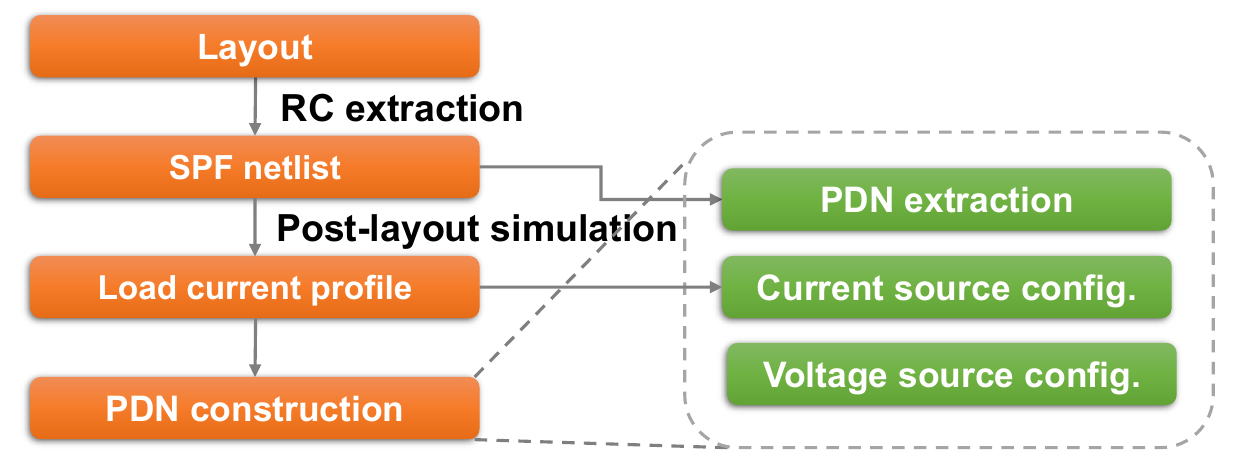}}
\caption{Workflow of generating PDN benchmark.}
\label{fig_wf}
\end{figure}

Fig. \ref{fig_wf} shows the workflow of generating the PDN benchmark. The post-layout netlist (SPF file) is first extracted from the GDS file using StarRC \cite{starrc}. The power delivery network including all parasitic capacitors and resistors is stripped off from the SPF netlist. Then we perform a post-layout simulation using FineSim \cite{finesim}, where the load current is measured at each power grid. The benchmark is generated by setting current sources to the measured values and voltage sources to the nominal operating voltage.

\subsection{Load Current Profiling and Modeling}
\begin{figure}[tb]
  \setlength{\abovecaptionskip}{0 cm}
\centerline{\includegraphics[width=1.0\linewidth]{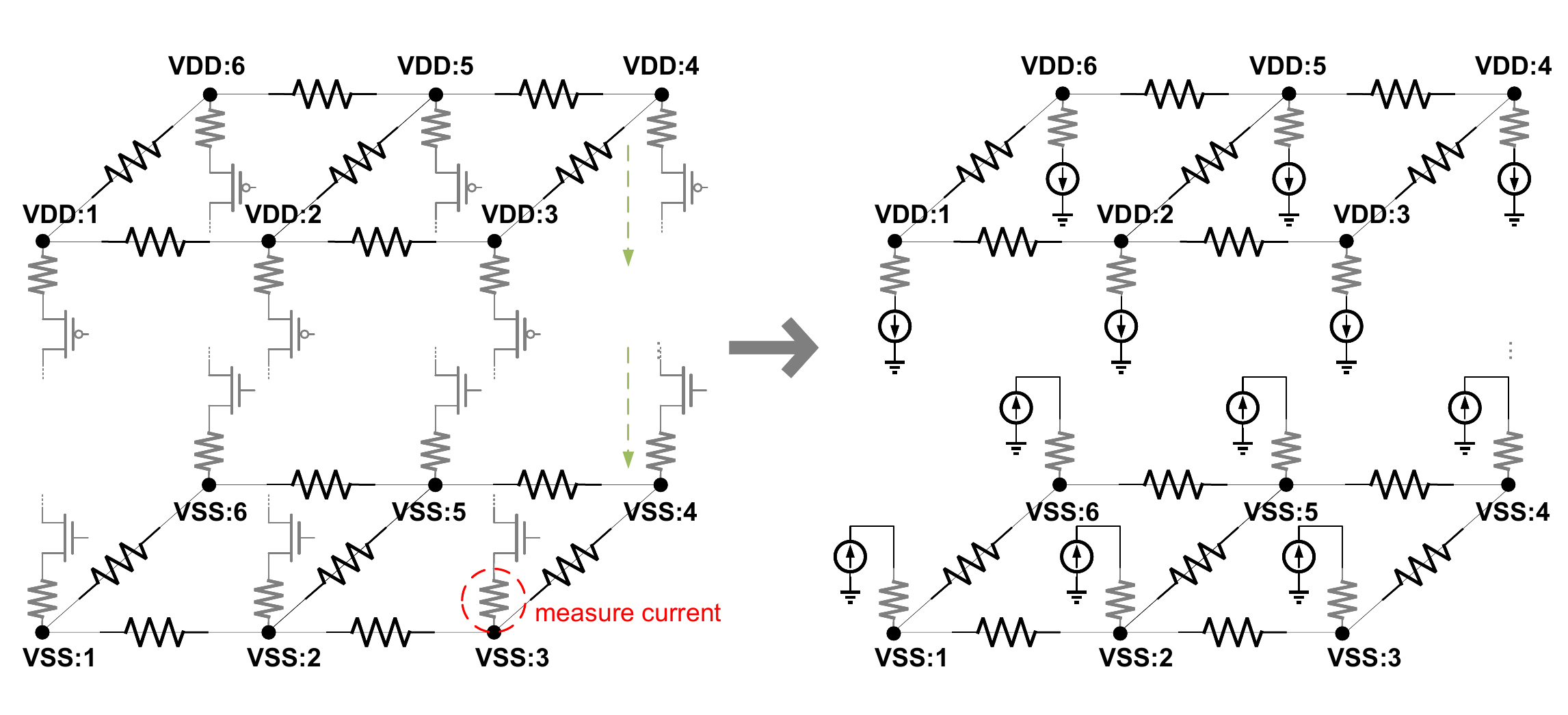}}
\caption{Illustration of a PDN and the load current profiling.}
\label{fig_2}
\end{figure}

To construct a more realistic PDN benchmark, the network should be extracted from a real circuit design. For an SRAM design, the power net may have different patterns on different metal layers and can be complicated when considering the via resistance and coupling capacitance. The proposed benchmarks are constructed based on the post-layout netlists after parasitic extraction \cite{starrc,wang2005improved} and remain a complete PDN topology. Fig. \ref{fig_2} shows an illustration of the PDN, where the power and ground networks are separated and consist of enormous parasitic resistors and capacitors (capacitors are not shown for simplicity). 

Terminal ports of devices connect the PDN through parasitic resistors. When the circuits are activated, the current is drawn from the power nodes and then flows into the ground net through stacked devices (sometimes the current can flow reversely).
We identify the power/ground grids in SPF file by matching the pattern where a parasitic resistor connects a VDD/VSS node and a terminal port of a circuit device, as shown in Fig. \ref{fig_2}. The load current can be obtained by measuring these resistors with the $.meas$ or $.probe$ statement in SPICE. The positive and negative terminals are probed for two-terminal devices including diodes, poly resistors, MOM capacitors. For a 4-port MOS, we only measure its source current and drain current since the magnitude of the gate or bulk current is much smaller than the channel current. 

\begin{figure}[tb]
  \setlength{\abovecaptionskip}{0 cm}
\centerline{\includegraphics[width=0.7\linewidth]{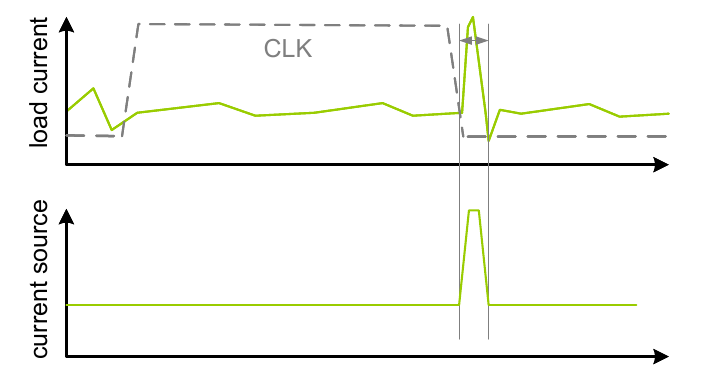}}
\caption{Load current modeling.}
\label{fig_3}
\vspace{-0.5em}
\end{figure}

Due to the existence of clock signals, the load current has an irregular shape, fluctuating between the minimum and maximum values, and the pattern is repeated in each cycle. Here we model load current as a narrow pulse, shown in Fig. \ref{fig_3}. Load current could be either a positive or a negative value that depends on whether it draws current from the PDN or injects current into the PDN. For a negative load current, the amplitude of the pulse matches the minimum value that we have measured in the simulation, and vice versa. The mean value of the current source will also be the same as the measured one. The rise/fall time, pulse width, and the period of the current source are set according to the simulated waveform. 

\subsection{Voltage Source Setting}
\begin{figure}[tb]
  \setlength{\abovecaptionskip}{0 cm}
\centerline{\includegraphics[width=1.0\linewidth]{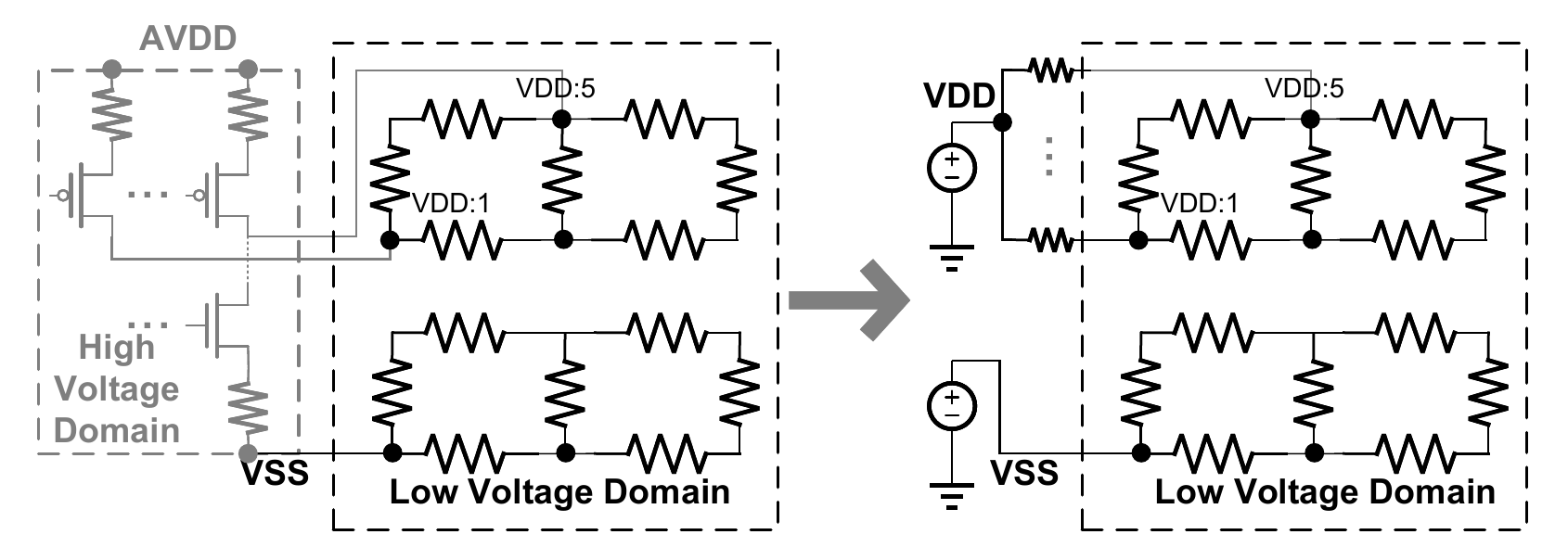}}
\caption{Processing the input power port.}
\label{fig_4}
\end{figure}

A design used in the proposed benchmark set (details in \ref{s3b}) has multiple supply voltages, usually a high-voltage domain and a low-voltage domain. Considering an open-sourced benchmark must protect the intelligence property and be unrelated to technology, we remove the DC-DC converts or header switches in designs. However, we still keep the high and low power networks intact. In the SPF file, the single input power port is split into multiple power nodes (Fig. \ref{fig_4}), connecting to driving transistors in the DC-DC converter or header switches. Here we merge these split nodes into a global port through small resistance resistors and attach a global voltage source to it. The voltage source is analog to the output of the DC-DC converter or header switches, which can feed enough current into the PDN. 

In SRAM-PG, all technology-related devices and other interconnections are filtered out. Power consumed by all circuit devices is converted to load current sources. The PDNs comprise voltage sources, current sources, resistors, and capacitors that include ground capacitance and coupling capacitance.

\section{Benchmark Description}
In this section, we introduce the four SRAM designs for generating SRAM-PG.

\subsection{SSRAM}
\begin{figure}[tb]
  \setlength{\abovecaptionskip}{0 cm}
\centerline{\includegraphics[width=1.0\linewidth]{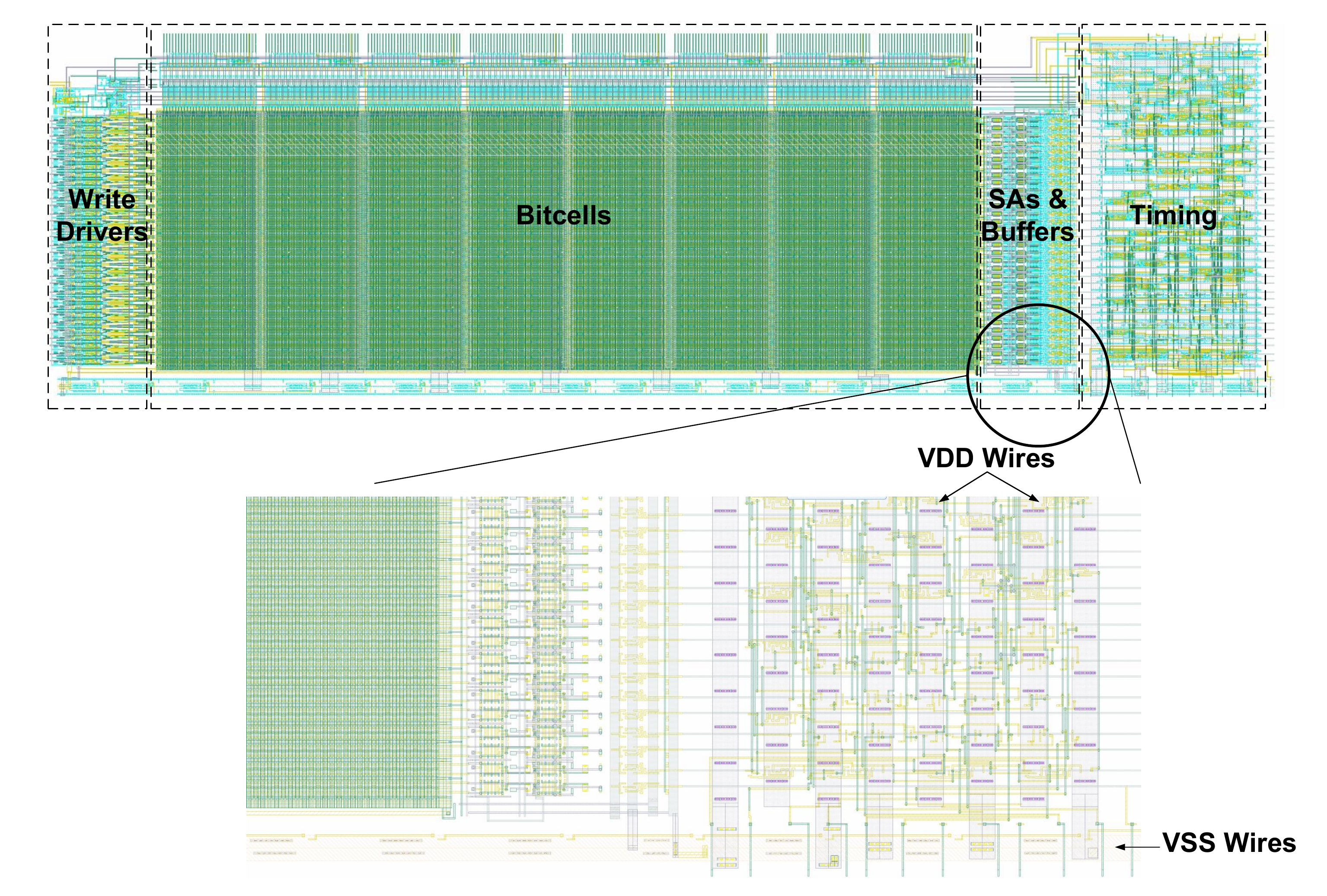}}
\caption{Layout of SSRAM macro.}
\label{fig_ssram}
\end{figure}

\begin{figure}[tb]
  \setlength{\abovecaptionskip}{0 cm}
\centerline{\includegraphics[width=0.8\linewidth]{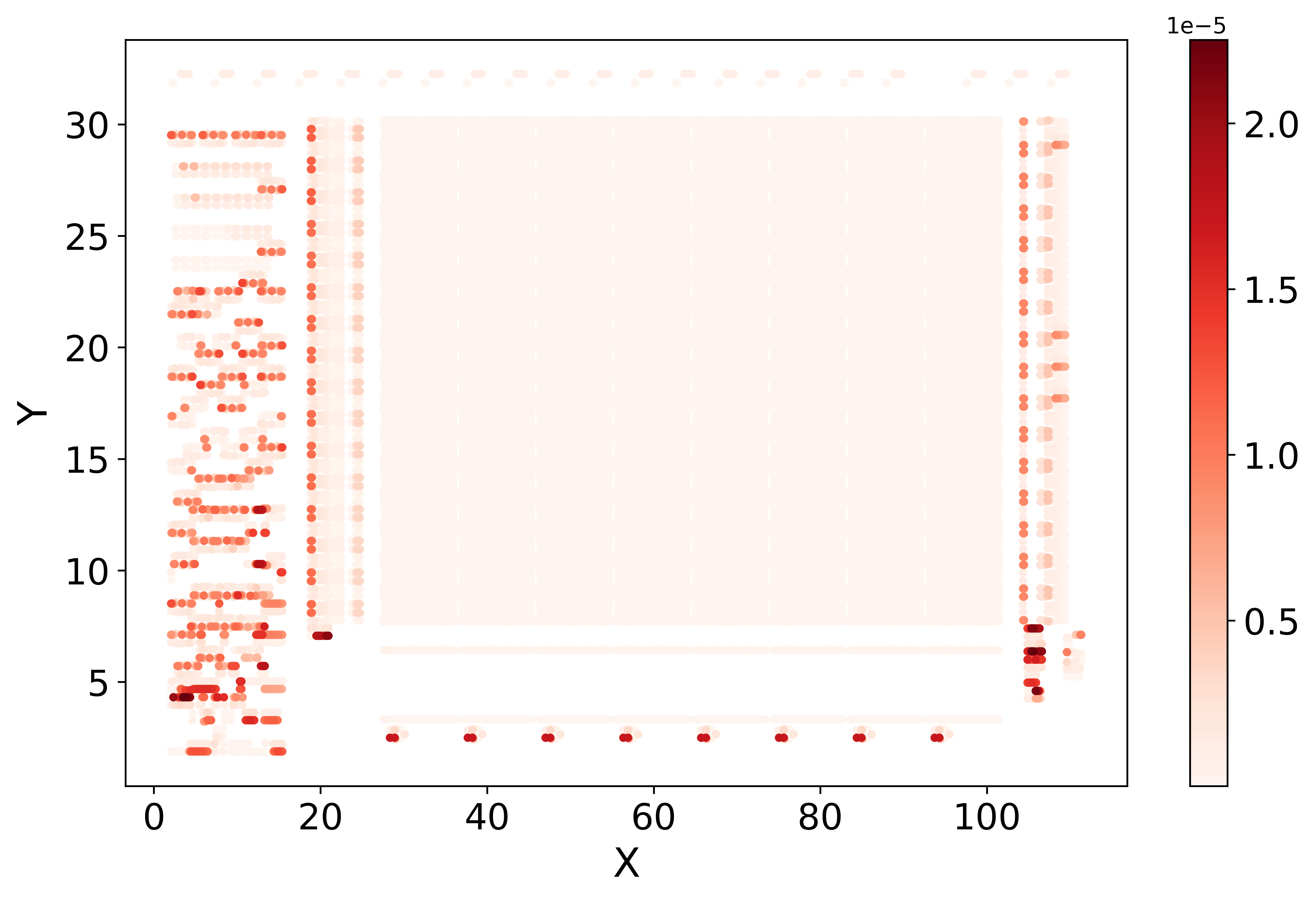}}
\caption{Current map of SSRAM macro.}
\label{fig_cm_ssram}
\end{figure}

SSRAM (Fig. \ref{fig_ssram}) is a macro of the basic SRAM array used by \cite{tscache}, which boosts the cache frequency and improves energy efficiency under low supply voltages. This design targets the near-threshold voltage domain and is validated through the tape-out measurement. The macro has 1KB capacity with 256 rows and 32 columns based on the medium density 6T SRAM cell provided in 28nm TSMC PDK. The PDNs of SSRAM are stacked from M1 to M6 and contain over 76K nodes. Main VDD and VSS wires are drawn using the M5 and M6 metal layers. They are around the boundary of the bitcell array and cover the timing module. The VDD and VSS networks have 20,602 and 20,278 load current sources respectively. The voltage source is set to 0.5V. Fig. \ref{fig_cm_ssram} shows the current map of SSRAM, where the peripheral circuits consume a large portion of power.

\subsection{Ultra8T SRAM}\label{s3b}
\begin{figure}[tb]
  \setlength{\abovecaptionskip}{0 cm}
\centerline{\includegraphics[width=1.0\linewidth]{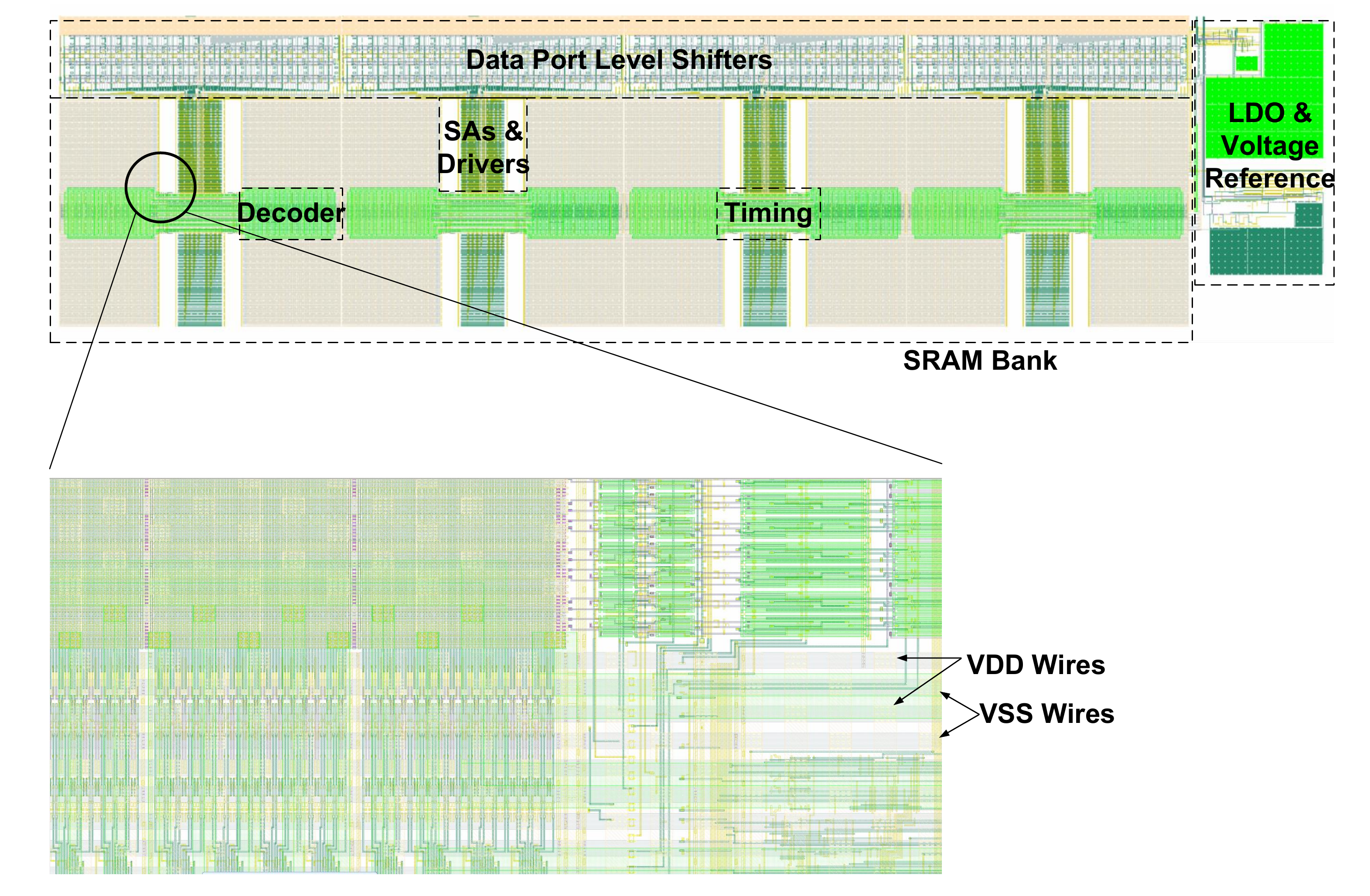}}
\caption{Layout of Ultra8T SRAM macro.}
\label{fig_u8t}
\end{figure}

\begin{figure}[htbp]
  \setlength{\abovecaptionskip}{0 cm}
\centerline{\includegraphics[width=0.8\linewidth]{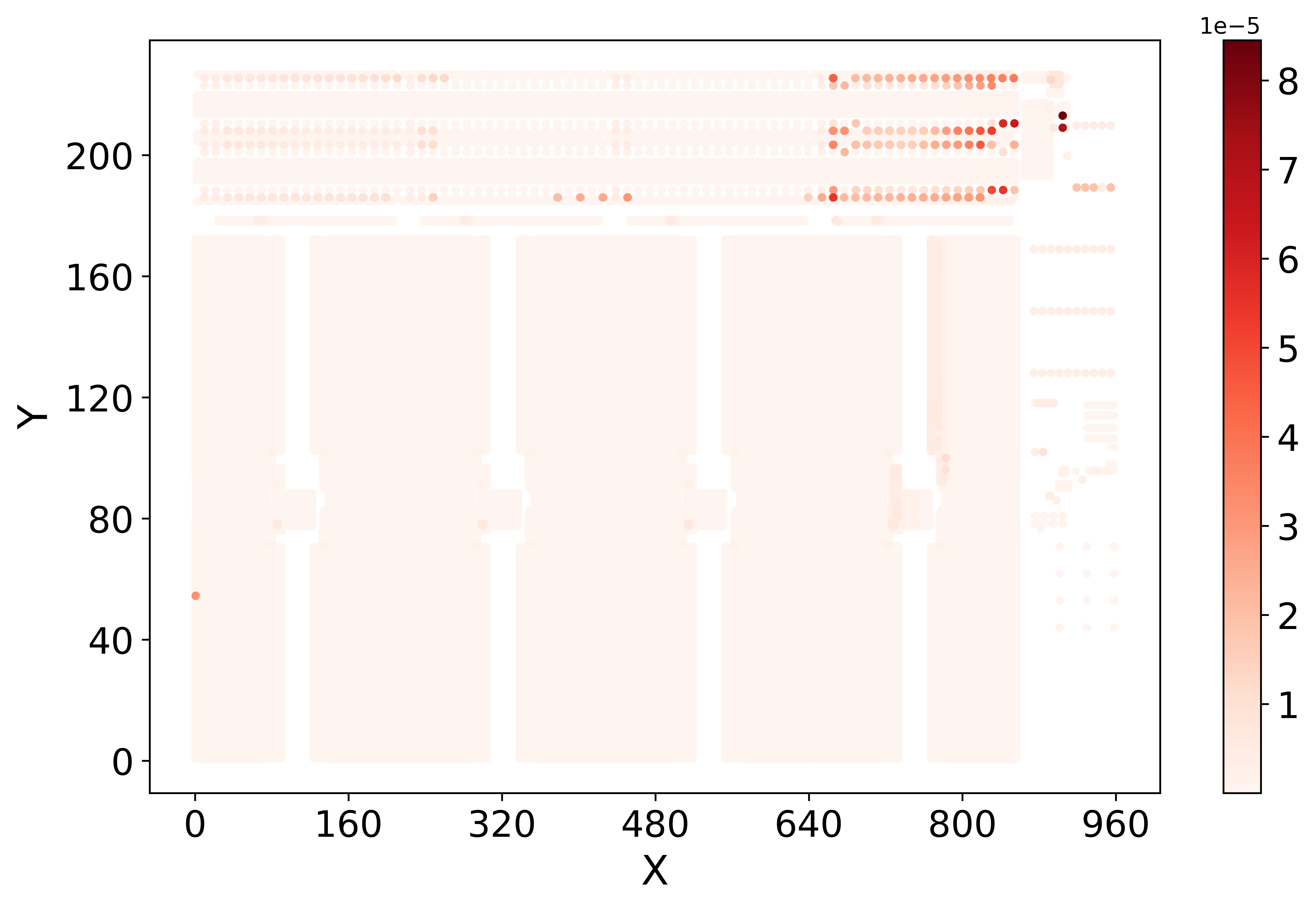}}
\caption{Current map of Ultra8T SRAM macro.}
\label{fig_cm_u8t}
\end{figure}

Ultra8T SRAM \cite{ultra8t} is a sub-threshold design that can aggressively reduce the operating voltage by using a leakage detection strategy without any additional hardware overhead (see Fig. \ref{fig_u8t}). It is based on 28nm TSMC CMOS technology and is comprised of an SRAM bank and analog circuits including level shifters, a low-dropout regulator (LDO), and voltage reference. Ultra8T has multiple voltage domains where the analog circuits are operated at 0.8V and the adjustable LDO provides 0.4V VDD for the SRAM bank. An SRAM bank consists of 4 sub-banks, and each sub-bank is formed by 4 basic bitcell arrays. The PDNs from this design are extracted from both the SRAM bank and analog parts. The PDNs include M1 to M7 layers and are drawn with high density and large wire width to avoid the large IR drop under low supply voltages. The voltage sources are set to 0.4V and 0.8V. The entire VDD and VSS networks contain 835,346 and 1,187,316 load current sources respectively. Fig. \ref{fig_cm_u8t} shows the current map, where level shifters consume more power than the memory core.

\subsection{Sandwich-RAM}\label{s3c}
\begin{figure}[tb]
  \setlength{\abovecaptionskip}{0 cm}
\centerline{\includegraphics[width=1.0\linewidth]{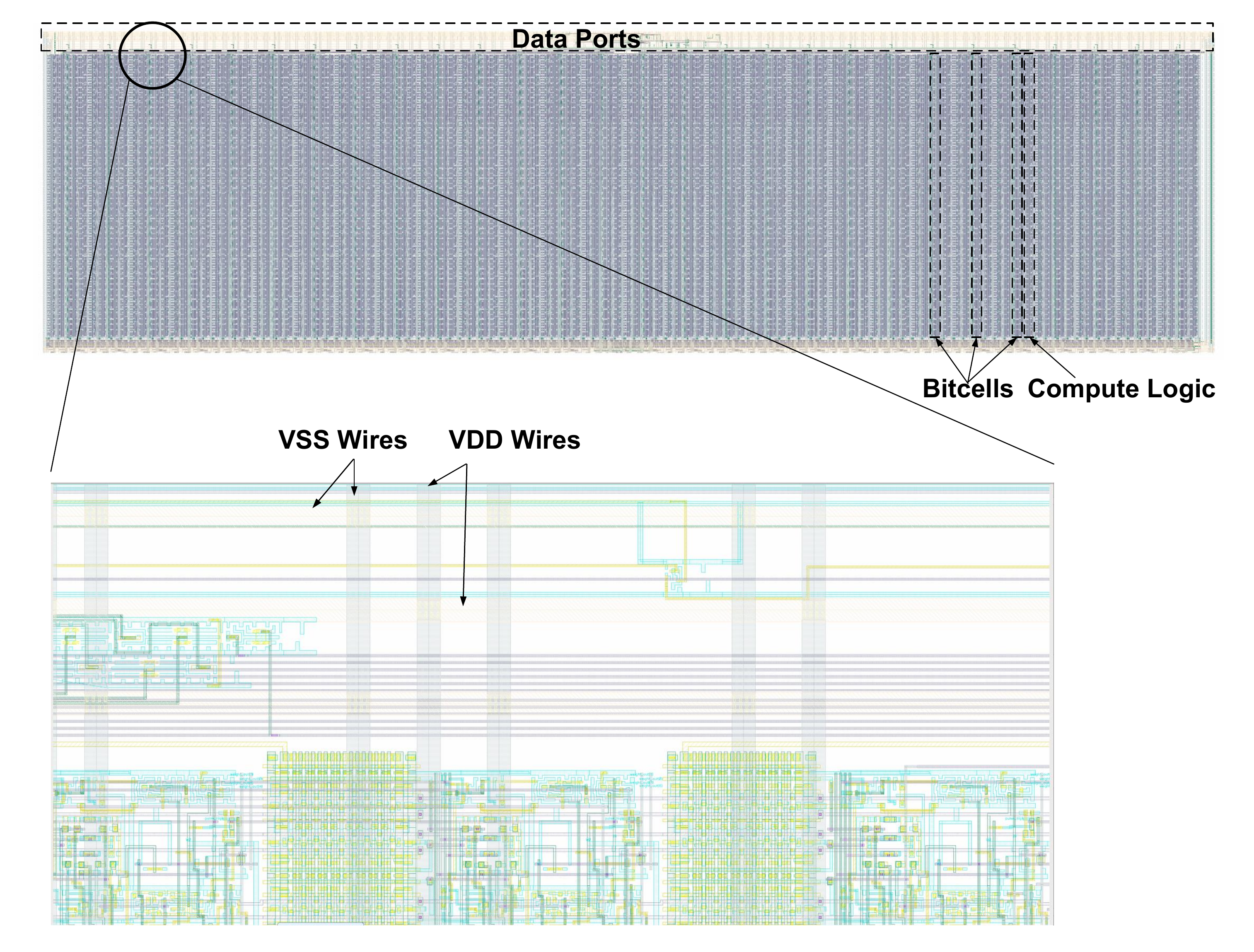}}
\caption{Layout of Sandwich-RAM macro.}
\label{fig_sandwich}
\end{figure}

\begin{figure}[htbp]
  \setlength{\abovecaptionskip}{0 cm}
\centerline{\includegraphics[width=0.8\linewidth]{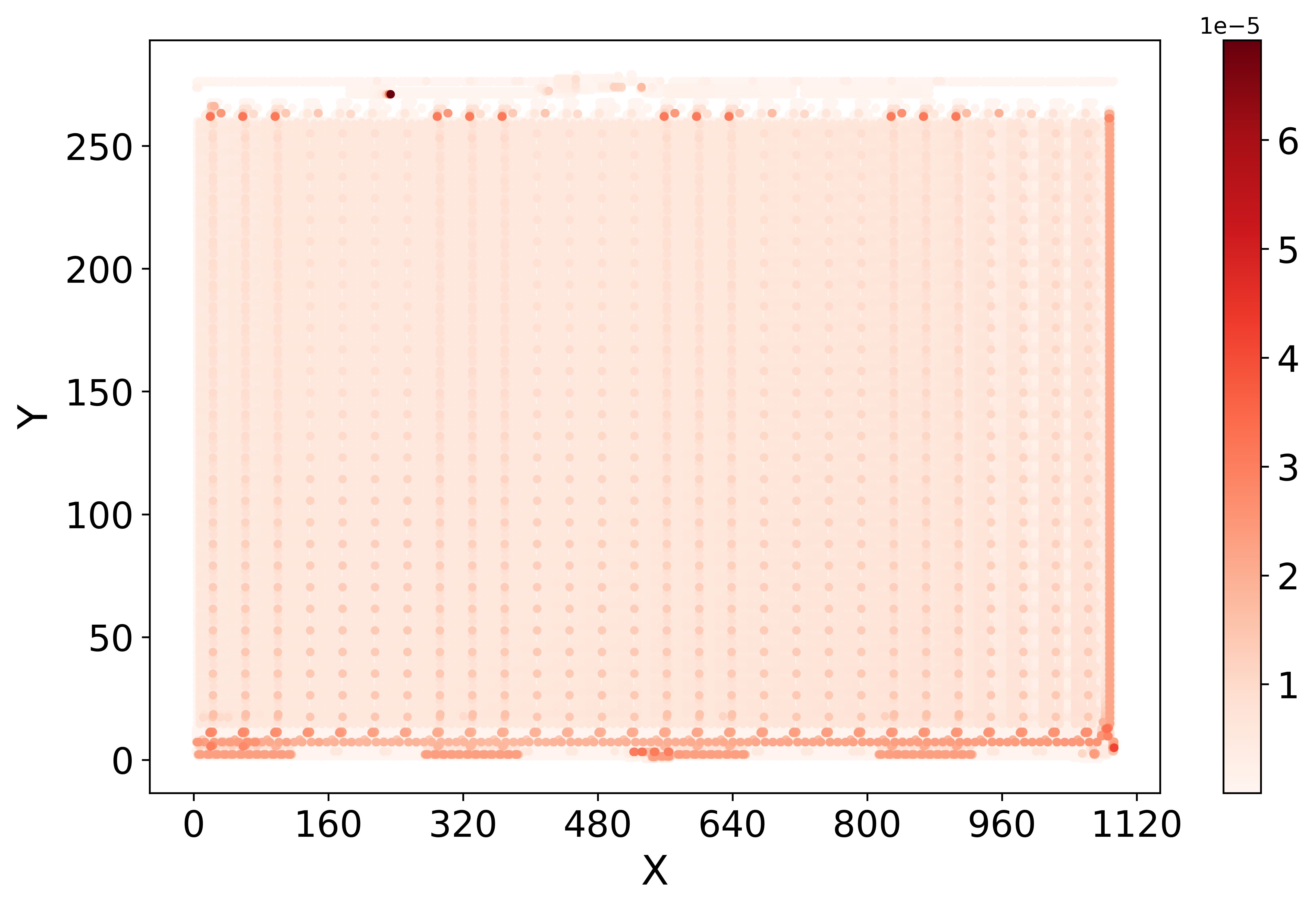}}
\caption{Current map of Sandwich-RAM macro.}
\label{fig_cm_sandwich}
\end{figure}

Sandwich-RAM \cite{sandwich} is an in-memory-computing design for binary weight convolutional neural networks that blends feature and partial-weight memory with a computing circuit together, like a sandwich, that achieves significantly fewer data access (see Fig. \ref{fig_sandwich}). It inserts a small and flexible reconfigurable analog-computation engine (based on pulse-width modulation) into the memory array.
Half of the circuits are the digital components for logic computing, while the other half contains SRAM arrays. The PDNs use layers from M1 to M6. The voltage source is set to 0.6V to achieve the highest energy efficiency. The VDD and VSS networks contain 987,966 and 1,309,162 load current sources respectively.   Fig. \ref{fig_cm_sandwich} shows the current map of Sandwich-RAM. Due to the in-memory-computing operations, the computing circuits and the memory core draw current from the PDN at the same time.

\subsection{SP8192W SRAM}
\begin{figure}[t]
  \setlength{\abovecaptionskip}{0 cm}
\centerline{\includegraphics[width=1.0\linewidth]{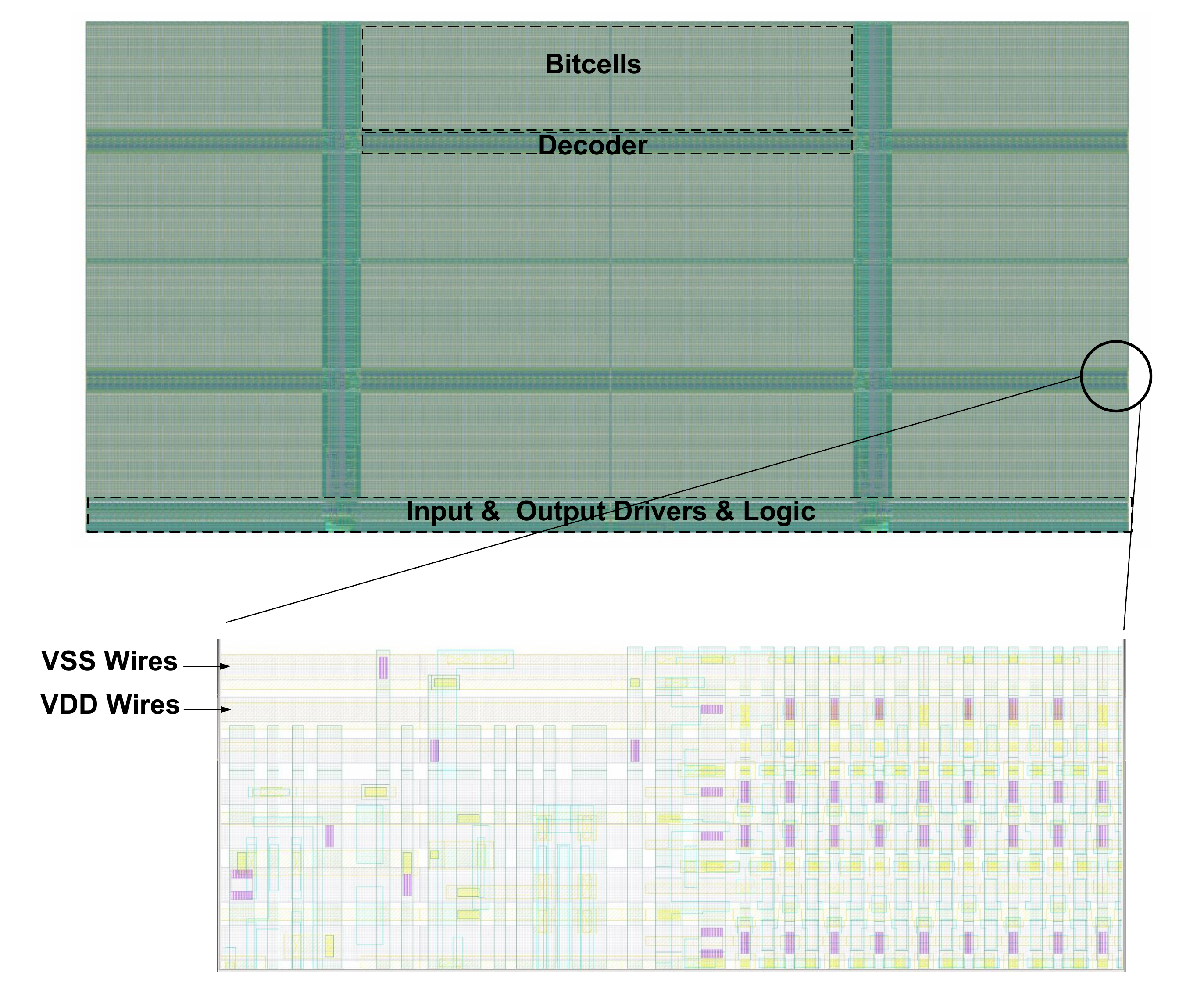}}
\caption{Layout of SP8192W SRAM macro.}
\label{fig_sp8192w}
\end{figure}

\begin{figure}[htbp]
  \setlength{\abovecaptionskip}{0 cm}
\centerline{\includegraphics[width=0.8\linewidth]{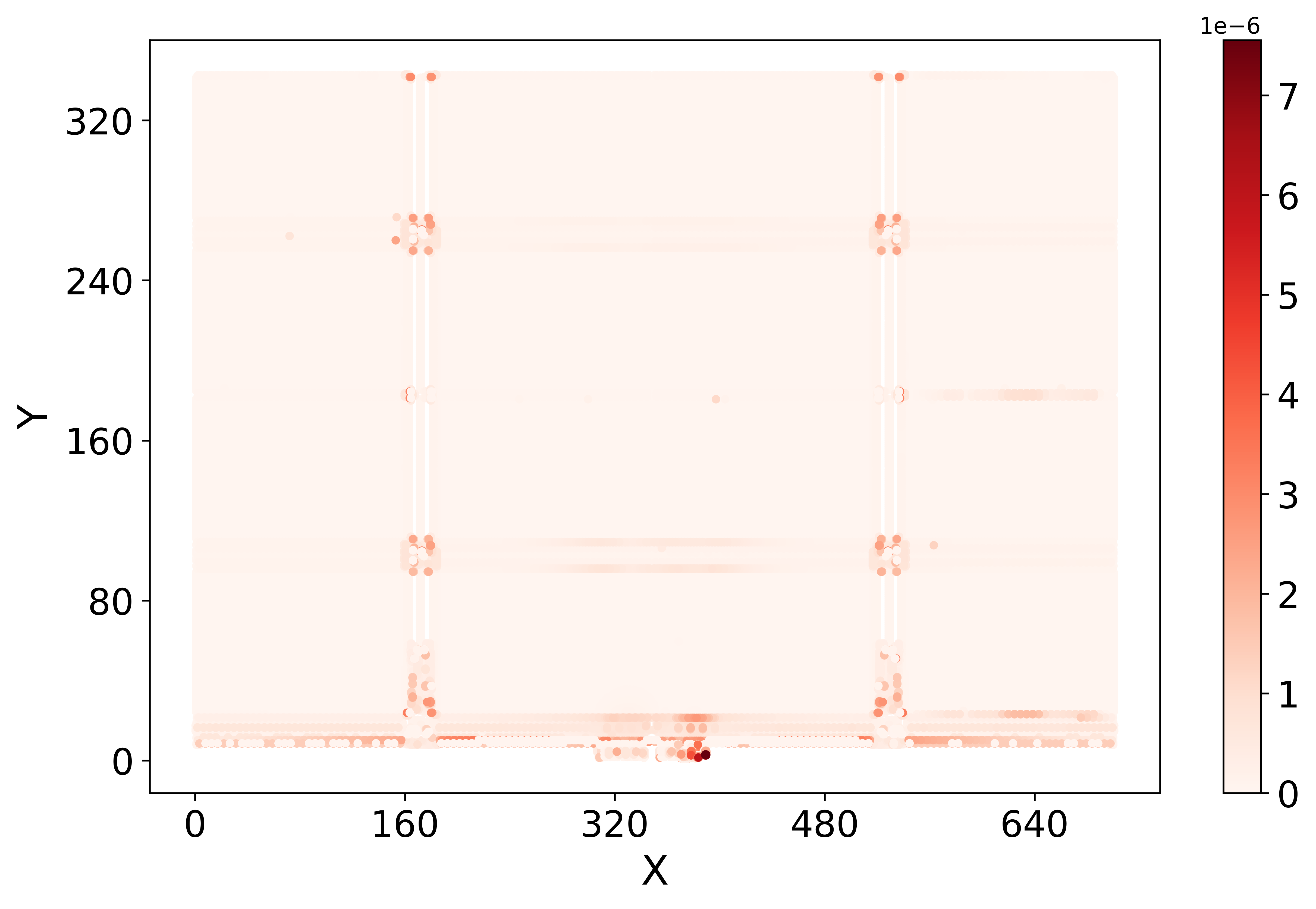}}
\caption{Current map of SP8192W SRAM macro.}
\label{fig_cm_sp}
\end{figure}

SP8192W SRAM (see Fig. \ref{fig_sp8192w}), which is the largest design in this work, is based on a single port 6T cell structure generated by the ARM SRAM compiler \cite{arm} with tremendously high density. This SRAM is fully compliant with the industry standard with good robustness. Memory cells consume over 90\% area of the design. This design uses high-threshold voltage transistors to reduce the leakage power. This leads to small voltage variations of PDN in the steady state of SRAM and the IR drop is also relatively small due to the high power wire density. Besides, SP8192W adopts a separate power supply for the memory core and the peripheral circuits, enabling the sleep mode to save energy. In the sleep mode, the peripheral power supply is turned off while the core power supply provides minimum power for the SRAM core to retain the data. The core and peripheral voltage sources are both set to 0.8V. The VDD and VSS networks contain 152,477 and 2,680,841 load current sources respectively. Fig. \ref{fig_cm_sp} shows the current map of the SRAM, where only a small portion of peripheral drivers are activated.

\begin{table}[hb]
  \setlength{\abovecaptionskip}{0 cm}
\begin{center}
\caption{Information of SRAM-PG}\label{tab_info}
\begin{threeparttable}
\begin{tabular}{@{}l@{~}|@{~}l@{~}|@{~}l@{~}@{~}l@{~}@{~}l@{~}@{~}l@{}}
\hline
 & Info. & SSRAM & Ultra8T* & Sandwich & SP8192W \\ \hline
 & Voltage (V) & 0.5 & 0.4, 0.8 & 0.6 & 0.8 \\
 & Area ($\mu m^2$) & 3,724 & 226,236.2 & 287,000 & 240,804.5 \\
General & Metal Layer & M1-M6 & M1-M7 & M1-M6 & M1-M4 \\
 & Node \# & 76,384 & 4,542,355 & 4,969,731 & 5,941,101 \\
 & Resistor \# & 106,796 & 7,103,220 & 7,243,343 & 10,050,878 \\
 & Capacitor \# & 37,506 & 1,783,923 & 4,427,494 & 2,472,252 \\ \hline
\multirow{2}{*}{VDD Net} & V Source \# & 1 & 2 & 1 & 2 \\
 & I Source \# & 20,602 & 835,346 & 987,966 & 152,477 \\ \hline
\multirow{2}{*}{VSS Net} & V Source \# & 1 & 1 & 1 & 1 \\
 & I Source \# & 20,278 & 1,187,316 & 1,309,162 & 2,680,841 \\ \hline 
\end{tabular}
    \begin{tablenotes}
        \footnotesize
        \item[*] Voltage sources are set to 0.4V and 0.8V for low and high voltage domains, respectively.
    \end{tablenotes}
\end{threeparttable}
\end{center}
\end{table}

Statistical information of SRAM-PG is listed in Table \ref{tab_info}, including the area, net, resistor, capacitor numbers, and node numbers in VDD and VSS networks. The smallest benchmark is SSRAM, which only contains 76 thousand net nodes, 106 thousand resistors, and 37 thousand capacitors. Ultra8T and Sandwich-RAM have a similar scale, including over 4 million nodes. Besides, Sandwich-RAM has 4.4 million capacitors, which is the largest number of capacitors among the 4 designs. The VSS network has more load current nodes than the VDD network in the last 3 benchmarks. The largest benchmark is SP8192W SRAM generated by the SRAM compiler, with nearly 6 million nodes, over 10 million resistors, and 2.4 million load current sources.

\section{Application Notes}
\begin{figure}[h]
  \setlength{\abovecaptionskip}{0 cm}
\centerline{\includegraphics[width=0.7\linewidth]{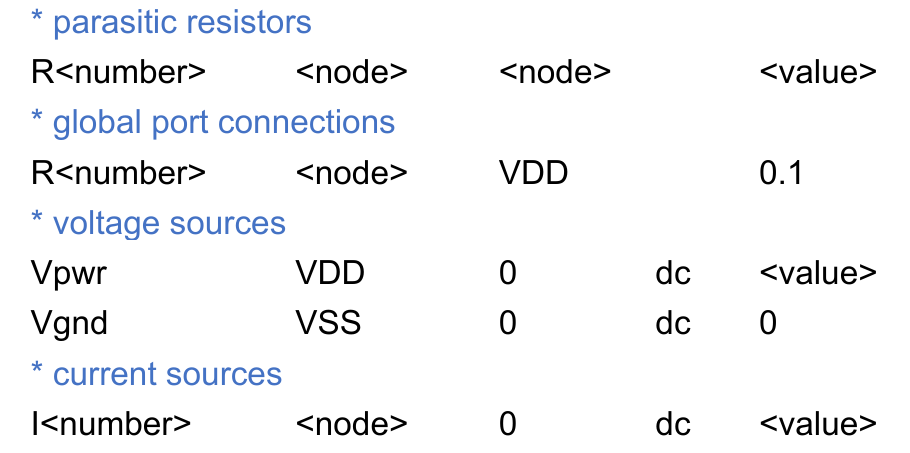}}
\caption{Template of SPICE file for DC analysis.}
\label{fig_dc}
\end{figure}

\begin{figure}[h]
  \setlength{\abovecaptionskip}{0 cm}
\centerline{\includegraphics[width=0.95\linewidth]{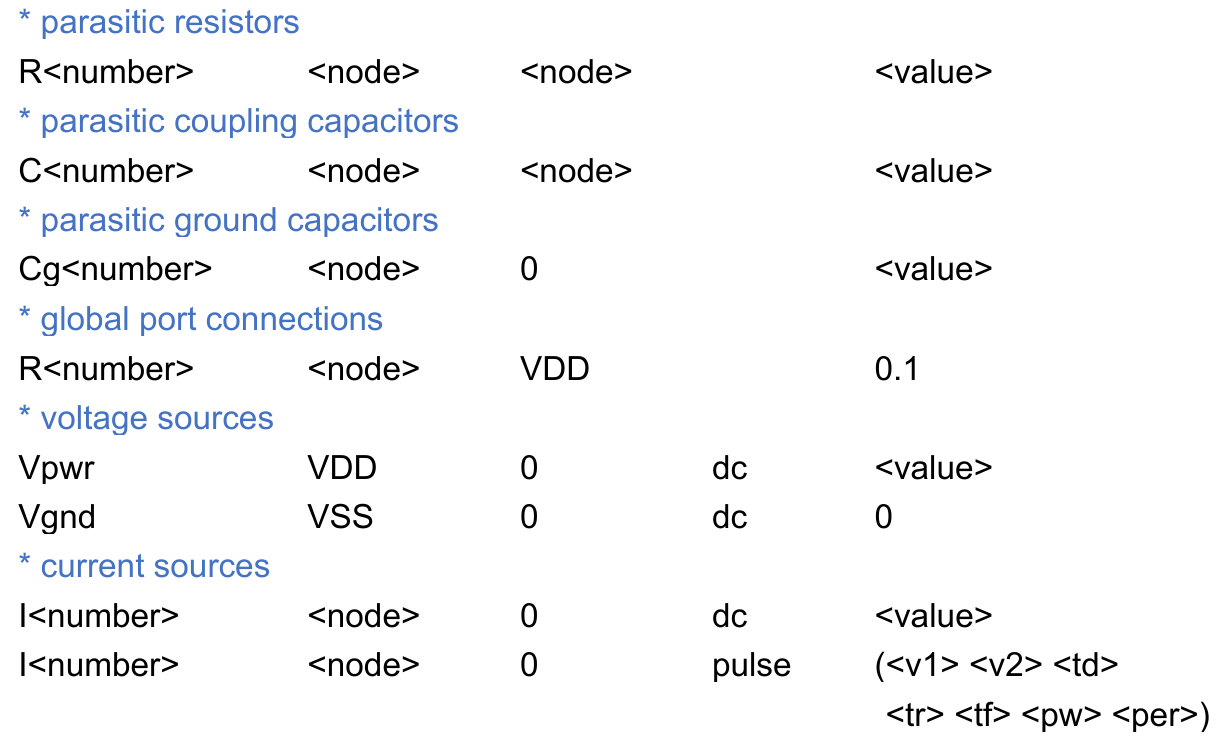}}
\caption{Template of SPICE file for transient analysis.}
\label{fig_trans}
\end{figure}

The proposed benchmarks support both DC and transient analysis. The corresponding files follow SPICE format, whose templates are shown in Fig. \ref{fig_dc} and Fig. \ref{fig_trans}. And the solution templates are also given. 

\subsubsection{design\_name\_dc.sp}
All extracted parasitics of PDNs are written into this file. It also contains all voltage/current sources, analysis statements, options, and variables to be probed or measured. The values of parasitic resistors are set according to the SPF file. The template of this file is shown as Fig. \ref{fig_dc}. The resistor connecting the net node to a global port exists only in designs with dc-dc converters or header switches. All voltage and current sources are DC sources. The naming rule for the power node and the ground node are VDD:$<$number$>$ and VSS:$<$number$>$, respectively.

\subsubsection{design\_name\_trans.sp} 
Fig. \ref{fig_trans} shows the transient file template. Different from the DC file, it has the ground parasitic capacitor and the coupling capacitors. The pulse current sources are provided only for a part of the power nodes (connecting activated circuit devices), where the parameters in the brackets denote the initial current value, maximum value, delay time, rise time, fall time, pulse width, and period. The simulation time is fixed at 40ns for all four benchmarks. %

\subsubsection{design\_name\_trans(dc).sol} 
This file lists the accurate solution for the benchmark. Each row is a key-value pair, $<node>$ $<value>$. The DC solution file contains all available nodes in the PDN. In the transient solution, a specific node name is first given, followed by key-value pairs, $<time>$ $<value>$, in several rows. To reduce the size of the solution file, we only store the transient solutions of 20 nodes that are randomly picked from the PDN. The key-value pairs are printed for each time step.


The maximum IR drops obtained in DC analysis and transient analysis for SRAM-PG are listed in Table \ref{tab_ir} (these results are simulated by FineSim, and may have a little variation using different circuit solvers or RC-reduction strategies).
For DC analysis, the minimum IR drop is 0.467mV for SSRAM, and the maximum value is 38.479mV for Sandwich-RAM. As for transient analysis, the load current pulses have a large impact on the PDN. The IR drop can increase by up to 9.57X compared to that in DC analysis. This phenomenon further demonstrates the necessity of transient analysis during PDN design and optimization.

\begin{table}[htb]
  \setlength{\abovecaptionskip}{0 cm}
\begin{center}
\caption{Maximum IR Drops of SRAM-PG in Transient and DC Analyses}\label{tab_ir}
\begin{threeparttable}
\begin{tabular}{@{~}l@{~}|l@{~}|l@{~}l@{~}l@{~}l@{~}}
\hline
Analysis & Networks & SSRAM & Ultra8T\tnote{*} & Sandwich & SP8192W \\ \hline
\multirow{2}{*}{.Tran} & VDD (mV) & 12.450 & 16.490, 50.050 & 53.020 & 1.980 \\
 & VSS (mV) & 4.029 & 52.723 & 81.154 & 1.109 \\ \hline
\multirow{2}{*}{.DC} & VDD (mV) & 1.300 & 6.050, 38.280 & 25.640 & 0.990 \\
 & VSS (mV) & 0.467 & 22.477 & 38.479 & 1.109 \\ \hline
\end{tabular}
    \begin{tablenotes}
        \footnotesize
        \item[*] Left number and right number are collected from 0.4V and 0.8V power networks, respectively.
    \end{tablenotes}
\end{threeparttable}
\end{center}
\end{table}

\section{Conclusions}
In this work, we develop and describe a new set of PDN benchmarks (named \textbf{SRAM-PG}) that are based on four real SRAM designs, and make them publicly available. Compared to existing PDN benchmarks, the proposed ones are comprised of full RC parasitics that are preserved after RC extraction. The load current is modeled to match the mean and maximum values collected from the post-layout simulations. In general, the proposed benchmarks reflect more up-to-date features of the PDNs in practical ICs. Therefore, they can be leveraged by research works that are in the fields of large-scale IR drop analysis, EM analysis, RC reduction, etc.

\bibliography{refs}
\bibliographystyle{IEEEtran}

\end{document}